\documentclass[aps,superscriptaddress,nofootinbib,12pt,tightenlines]{revtex4}

\makeatletter

\usepackage{natbib}
\usepackage{graphicx}
\usepackage{dcolumn}
\usepackage{amsmath}
\begin{document}
\title{An empirical behavioral model of price formation}

\author{Szabolcs Mike}
\affiliation{Santa Fe Institute, 1399 Hyde Park Road, Santa Fe, NM 87501}
\affiliation{Budapest University of Technology and Economics,
H-1111 Budapest, Budafoki \'ut 8, Hungary}


\author{J. Doyne Farmer}
\affiliation{Santa Fe Institute, 1399 Hyde Park Road, Santa Fe, NM 87501}


\begin{abstract}
Although behavioral economics has demonstrated that there are many situations where rational choice is a poor empirical model, it has so far failed to provide quantitative models of economic problems such as price formation.  We make a step in this direction by developing empirical models that capture behavioral regularities in trading order placement and cancellation using data from the London Stock Exchange.  For order placement we show that the probability of placing an order at a given price is well approximated by a Student distribution with less than two degrees of freedom, centered on the best quoted price.  This result is surprising because it implies that trading order placement is symmetric, independent of the bid-ask spread, and the same for buying and selling.   We also develop a crude but simple cancellation model that depends on the position of an order relative to the best price and the imbalance between buying and selling orders in the limit order book.  These results are combined to construct a stochastic representative agent model, in which the orders and cancellations are described in terms of conditional probability distributions.  This model is used to simulate price formation and the results are compared to real data from the London Stock Exchange.  Without adjusting any parameters based on price data, the model produces good predictions for the magnitude and functional form of the distribution of returns and the bid-ask spread.
\end{abstract}
\maketitle

\tableofcontents

\section{Introduction}

In the last two decades the field of behavioral finance has presented an increasingly large number of examples where equilibrium rational choice models are a poor description of real economic behavior\footnote{This may be partly because of other strong assumptions that typically accompany such models, such as complete markets.  Until we have predictive models that drop these assumptions, however, we will not know whether more realistic assumptions in rational choice models are sufficient to solve these problems.} (Hirschleifer \citeyear{Hirschleifer01}, Barberis and Thaler \citeyear{Barberis03}, Camerer \citeyear{Camerer03}, Thaler \citeyear{Thaler05}).  While this evidence may be compelling, so far behavioral finance has largely been a negative science.  It illustrates problems with the mainstream paradigm, but it fails to provide a positive alternative with quantitative predictive power\footnote{There are several examples of models that make {\it qualitative} predictions, such as in the popular book by Schleifer (\citeyear{Schleifer00}).}. There are many efforts underway to build a better foundation for economics based on psychological evidence, but this imposes a difficult hurdle for building quantitative theories.  The human brain is a complex and subtle instrument, and the distance from psychology to prices is large.

In this study we take advantage of the fact that electronic markets provide a superb laboratory for studying patterns in human behavior.   Market participants make decisions in an extremely complex environment, but in the end these decisions are reduced to the simple actions of placing and canceling trading orders.  The data that we study here contains hundreds of millions of records of both trading orders and prices, allowing us to reconstruct the state of the market at any instant in time.  

Our goal here is to capture behavioral regularities in order placement and cancellation, and to exploit these regularities to achieve a better understanding of price formation.  The practical component of this goal is to understand statistical properties of prices, such as the distribution of price returns and the bid-ask spread.  Serious interest in the distribution of prices began with Mandelbrot's (\citeyear{Mandelbrot63}) study of cotton prices, in which he showed that logarithmic price returns are far from normal, and suggested that they might be drawn from a Levy distribution.   There have been many studies since then, most of which indicate that the cumulative distribution of logarithmic price changes has tails that asymptotically scale for large $| r |$ as a power law of the form $|r|^{-\alpha}$, where $r(t) = \log p(t) - p(t - \tau)$ is the logarithmic return at time $t$ on time scale $\tau$ and $p$ is the price (Fama \citeyear{Fama65},  Officer \citeyear{Officer72}, Akgiray, Booth and Loistl \citeyear{Akgiray89}, Koedijk, Schafgans and de Vries \citeyear{Koedijk90},   Loretan \citeyear{Loretan94}, Mantegna and Stanley \citeyear{Mantegna95},  Longin \citeyear{Longin96},  Lux \citeyear{Lux96},  Muller, Dacorogna and Pictet \citeyear{Muller98}, Plerou et al. \citeyear{Plerou99}, Rachev and Mittnik \citeyear{Rachev00}, Goldstein, Morris and Yen \citeyear{Goldstein04}).  The exponent $\alpha$, which takes on typical values in the range $2 < \alpha < 4$, is called the {\it tail exponent}.  It is important because it characterizes the risk of extreme price movements and corresponds to the threshold above which the moments of the distribution become infinite.  Having a good characterization of price returns has important practical consequences for risk control and option pricing.

From a theoretical point of view, the heavy tails of price returns excite interest among physicists because they suggest nonequilibrium behavior.  A fundamental result in statistical mechanics is that, except for unusual situations such as phase transitions, equilibrium distributions are either exponential or normal distributions\footnote{For example, at equilibrium the distribution of energies is exponentially distributed and the distribution of particle velocities is normally distributed.  This is violated only at phase transactions, e.g. at the transition between a liquid and a gas.}.  The fact that price returns have tails that are heavier than this suggests that markets are not at equilibrium.  Although the notion of equilibrium as it is used in physics is very different from that in economics, the two have enough in common to make this at least an intriguing possibility.  Many models have been proposed that attempt to explain the heavy tails of price returns (Arthur et al. \citeyear{Arthur97}, Bak, Pacuski and Shubik \citeyear{Bak97}, Brock and Hommes \citeyear{Brock99}, Lux and Marchesi \citeyear{Lux99},  Chang, Stauffer and Pandey \citeyear{Chang02}, LeBaron \citeyear{LeBaron03}, Giardina and Bouchaud \citeyear{Giardina03}, Gabaix et al. \citeyear{Gabaix03}, Challet, Marsili and Zhang \citeyear{Challet05}).  These models have a wide range in the specificity of their predictions, from those that simply demonstrate heavy tails to those that make a more quantitative prediction, for example about the tail exponent $\alpha$.  However, none of these models produce quantitative predictions of the magnitude and functional form of the full return distribution.

The bid-ask spread $s$ is another important market property.  It can be defined as $s(t) = \pi_a(t) -  \pi_b(t)$, where $\pi_a(t)$ is the logarithm of the best selling price offered at time $t$ and $\pi_b(t)$ is the logarithm of the best buying price.  The spread is important as a benchmark for transaction costs.  A small market order to buy will execute at the best selling price, and an order to sell will execute at the best buying price, so someone who first buys and then sells in close succession will pay the spread $s(t)$.  There is a substantial empirical and theoretical literature on the spread (a small sample of examples are Demsetz \citeyear{Demsetz68}, Stoll \citeyear{Stoll78}, Glosten \citeyear{Glosten88}, Glosten \citeyear{Glosten92}, Easley and O'Hara \citeyear{Easley92}, Foucault, Kadan and Kandel \citeyear{Foucault01}, Sandas \citeyear{Sandas01}), but all this work has a focus that is substantially different from ours here.

The model we develop here is a statistical description of the placement and cancellation of trading orders under a continuous double auction, which can be used to understand the statistical properties of price returns and spreads.  This model follows in the footsteps of a long list of other models that have tried to describe order placement as a statistical process (Mendelson
\citeyear{Mendelson82}, Cohen {\it et al.} \citeyear{Cohen85}, Domowitz and Wang \citeyear{Domowitz94},  Bollerslev, Domowitz and Wang \citeyear{Bollerslev97}, Bak {\it et al.} \citeyear{Bak97},  Eliezer and Kogan \citeyear{Eliezer98}, Tang \citeyear{Tang99}, Maslov \citeyear{Maslov00}, Slanina \citeyear{Slanina01}, Challet and Stinchcombe \citeyear{Challet01}, Daniels et al. \citeyear{Daniels03}, Chiarella and Iori, \citeyear{Chiarella02}, Bouchaud, Mezard and Potters \citeyear{Bouchaud02},  Smith et al. \citeyear{Smith03}).  For a more detailed narrative of the history of this line of work, see Smith et al. (\citeyear{Smith03}).   We build on the model of Daniels et al.  (\citeyear{Daniels03}), 
which assumes that limit orders, market orders, and cancellations can be described as independent Poisson processes, in which buying and selling have the same parameters.  Because it assumes that order placement is random except for a few constraints, it can be regarded as a zero intelligence model of agent behavior.  Although highly unrealistic in many respects, the zero intelligence model does a reasonable job of capturing the dynamic feedback and interaction between order placement on one hand and price formation on the other.  It predicts scaling laws for the volatility of returns and for the spread, which can be regarded as equations of state relating the properties of order flows to those of prices.  Farmer, Patelli and Zovko (\citeyear{Farmer05}) tested these predictions against real data from the London Stock Exchange and showed that, even though the model does not predict the absolute magnitude of these effects, it does a good job of capturing how average volatility and spread vary with changes in order flow. 

Despite these successes the zero intelligence model is inadequate in many respects.  Because of the unrealistic assumptions that order placement and cancellation are uniform along the price axis, to make comparisons with real data it is necessary to introduce an arbitrary interval over which order flow and cancellation densities are measured, and to assume that they vanish outside this interval.   This assumption introduces arbitrariness into the scale predictions.  In addition it produces price returns with non-white autocorrelations and a thin-tailed distribution, which do not match the data.

The model here has the same basic elements as the zero intelligence model, but each element is modified based on empirical analysis.  In the course of doing this we uncover regularities in order placement and cancellation that are interesting for their own sake.  The strategic motivation behind that patterns that we observe is in many cases not obvious -- it is not clear whether they are driven by rational equilibrium or irrational behavior.  We do not attempt to address this question here.  Instead we work in the other direction and construct a model of price formation.  The resulting model makes good predictions about the magnitude and the functional form of the distribution of returns and spreads.  The predictions are particularly good in the body of the distributions and not quite as good in the tails.  The model predicts a power law for the tails, though with tail exponents that are slightly larger than those of the real data.   We believe this is due to inadequacies in our cancellation model, which at this stage is still crude.

The paper is organized as follows:  Section \ref{data} discusses the market structure and the data set.  In Section~\ref{orderPlacement} we study the distribution of order placement conditioned on the spread and in Section~\ref{orderSigns} we discuss our approach to modeling the long-memory order signs (whether orders are to buy or to sell).  In Section~\ref{orderCancellation} we develop a model for order cancellation, and in Section~\ref{priceFormation} we develop a simulation based on the models of order flow that allows us to make predictions about returns and spreads.  Finally in the last section we summarize and discuss the implications and future directions of this work. 

\section{The market and the data\label{data}}

This study is based on data from the on-book market in the London Stock exchange.  These data contain all order placements and cancellations, making it possible to reconstruct the limit order book at any point in time.  In 1997 $57\%$ of the transactions in the LSE occurred in the on-book market and by 2002 this rose to $62\%$.  The remaining portion of the trading takes place in the off-book market, where trades are arranged bilaterally by telephone.   Off-book trades are published, but only after they have been arranged and have effectively already taken place.  Because the on-book market is public and the off-book market is not, it is generally believed that the on-book market plays the dominant role in price formation.  We will not use any information from the off-book market here.  For a more extensive discussion of the LSE market structure, together with some comparative analysis of the two markets, see Lillo, Mike and Farmer (\citeyear{Lillo05}).

The LSE on-book market is purely electronic.  There are no designated market makers (though there is no restriction on simultaneously placing orders to buy and to sell at the same time).  Timestamps are accurate to the second and sequencing is more accurate than that, though not always perfect.   Because we have a complete record we know whether orders are to buy or to sell, and we also know all transactions in both markets.  

We study data for three stocks, Astrazeneca (AZN), Lloyds (LLOY) and BHP Billiton (BLT).  The data we analyze are for the period from May 2000 - December 2002.   Counting an event as either an order placement or an order cancellation, there are $4.2$ million events for AZN, $652$ thousand of which are transactions; $3.4$ million for LLOY, $723$ thousand of which are transactions, and $1.7$ million events for BLT,  $297$ thousand of which are transactions.

During the trading day the LSE is a continuous double auction.  Trading begins with an opening auction and ends with a closing auction.  To keep things simple we remove these, and also remove the first hour and last half hour of trading data, i.e. we consider only data from 9:00 am to 4:00 pm.  We do this because near the auctions there are transient behaviors, such as the number of orders in the book building up and winding down, caused by the fact that many traders close out their books and the end of the day.  (Even so, this does not seem to be a large effect and does not make a great difference in our results). We paste together data from different days by freezing time outside of 9:00 - 4:00 on trading days.  In our data analyses we are careful not to include any price movements that span the daily boundaries.  This allows us to treat the data as if it were one continuously running market.

There are several different types of possible trading orders in the LSE.   The details are not important here.  We will simply call any order with a limit price attached a limit order, and any order that generates an unconditional execution a market order.   For convenience we will define an {\it effective market order} as any trading order that generates an immediate transaction, and an {\it effective limit order} as any order that does not.  The limit order book refers to the queue that holds limit orders waiting to be executed.  The priority for executing limit orders depends both on their price and on the time when they are placed, in the obvious way. 

\section{Order placement \label{orderPlacement}}

Even a brief glance at the data makes it clear that the probability for order placement depends on the distance from the current best prices.  This was studied in the Paris Stock Exchange by Bouchaud, Mezard and Potters (\citeyear{Bouchaud02}) and in the London Stock Exchange by Zovko and Farmer (\citeyear{Zovko02}).  Both groups studied only orders placed inside the limit order book.  They found that the probability for order placement drops off asymptotically as a power law of the form $x^{-\alpha}$.  The value of $\alpha$ varies from stock to stock, but is roughly $\alpha \approx 0.8$ in the Paris Stock Exchange and $\alpha \approx 1.5$ in the London Stock Exchange.  This means that in Paris the mean of the distribution does not exist and in London the second moment does not exist.  The small values of $\alpha$ are surprising because they imply a non-vanishing probability for order placement extremely far from the current best prices, where it would seem that the probability of ever making a transaction is exceedingly low.  Such orders are often replaced when they expire, which taken together with the fact that they their distribution lies on a continuous scaling curve with orders that are close to the best prices suggests that they are intentional.

Here we add to this earlier work by studying the probability of order placement inside the spread and the frequency of transactions conditional on the spread.  We will say that a new order is placed {\it inside the book} if its logarithmic limit price $\pi$ places it within the existing orders, i.e. so that for a buy order $\pi < \pi_b$ and for a sell order $\pi > \pi_a$.   We will say it is {\it inside the spread} if its limit price is between the best price to buy and the best price to sell, i.e. $\pi_b < \pi < \pi_a$.  Similarly, if it is a buy order it generates a transaction for $\pi \ge \pi_a$ and if it is a sell order for $\pi \le \pi_b$.  To simplify nomenclature, when we are speaking of buy orders, we will refer to $\pi_b$ as the {\it same best} price and $\pi_a$ as the {\it opposite best} price, and vice versa when we are speaking of sell orders.  We will define $x$ as the logarithmic distance from the same best price, with $x = \pi - \pi_b$ for buy orders and $x = \pi_a - \pi$ for sell orders.  Thus by definition $x = 0$ for the same best price, $x > 0$ for aggressive orders placed outside the limit order book, and $x < 0$ for less aggressive orders placed inside the limit order book. 

In deciding where to place an order, a trader needs to make a strategic trade off between certainty of execution on one hand and price improvement on the other.  One would naturally expect that for strategic reasons the limit prices of orders placed inside the book should have a qualitatively different distribution than those placed inside the spread.  To see why we say this consider a buy order.  If the trader is patient she will choose $\pi < \pi_b$.   In this case the order will sit inside the limit book and will not be executed until all buy orders with price greater than $\pi$ have been removed.  The proper strategic trade off between certainty of execution and price improvement depends on the position of all the other orders -- to gain more price improvement means more other orders have to be executed, which lowers the probability of execution.  In the limit where $\pi \ll \pi_b$ and there are many orders in the queue the execution probability and price improvement vary in a quasi-continuous manner with $\pi$.

The situation is different for an impatient trader.  Such a trader will choose $\pi > \pi_b$.  If she is very impatient and is willing to pay a high price she will choose $\pi \ge \pi_a$, which will result in an immediate transaction.  If she is of intermediate patience, she will place her order inside the spread.  In this case the obvious strategy is to place the order one price tick above $\pi_b$, as this is the best possible price with higher priority than any existing orders.  It would seem foolish to place an order anywhere else inside the spread\footnote{This reasoning neglects the consequences of time priority; as we will discuss later, when time priority is taken into account other values may be reasonable.}, as this gives a higher price with no improvement in priority of execution.  One would therefore naively expect to find that order placement of buy orders inside the spread is highly concentrated one tick above the current best price.  This suggests that as one moves away from the same best price, how new trading orders are placed inside the spread should be different from how they are placed inside the book.  The resulting distribution should be asymmetric around $\pi_b$. 

To model order placement we seek a good functional form for $p(x | s)$, the probability density for $x$ conditioned on the spread.  This is complicated by the fact that for an order that generates an immediate transaction the relative price $x$ is not meaningful.  Such an order can either be placed as a limit order with $x > s$ or as a market order, which has an effective price $x = \infty$.  In Farmer et al. (\citeyear{Farmer04}) we showed that for the LSE it is very rare for a market order to penetrate deeper than the opposite best price.  This restriction can be achieved either through the choice of limit price or by the choice of order size.  Thus two orders with very different limit prices may be equivalent from a functional point of view.  We resolve this ambiguity by lumping all orders with $x > s$ together and characterizing them by $P_\theta$, the probability that a trading order causes an immediate transaction\footnote{If only part of an order causes an immediate transaction we will treat it as two orders, one of which causes a transaction and one of which doesn't.}.  We thus restrict our attention to the probability $p(x | s)$ only for orders with $x < s$ and the probability $P_\theta$ for orders that generate transactions (which can be thought of as all orders with $x \ge s$).  We expect $P_\theta$ to depend on the spread.

After considerable exploratory analysis we have been led to make a simple hypothesis about order placement.    Our hypothesis is that
 \begin{eqnarray}
 \label{hyp1}
 p(x | s) & = & p(x),  -\infty < x < s,\\
 \label{hyp2}
 P_\theta (s) & = & \int_s^{\infty} p(x) dx,
 \end{eqnarray}
where $p(x)$ is a probability density function that is symmetric about $x = 0$ and that is well approximated by a Student distribution.  Equation (1) describes the prices where limit orders are placed, and equation (2) describes the probability of a transaction.  If the relative price $x$ drawn from $p(x)$ satisfies $-\infty < x < s$ then it is an effective limit order at price $\pi = \pi_b + x$, and if $x \ge s$ it is an effective market order.  Note that the symmetry of the hypothesized functional form contradicts the strategic reasoning given earlier.

The first hypothesis implies that $p(x, s) = p(x) p(s)$ on $-\infty < x < s$, i.e. that $x$ and $s$ are independent in this range.  We must be careful, however, since the range depends on $s$, so $p(x | s)$ is not fully independent of $s$.  This makes it useful to rewrite this hypothesis in a form that makes the dependence on $s$ more explicit.  As a convenient conceptual device, we can treat any effective market order to buy as if it were a limit order with a limit price $\pi = \pi_0$, where $\pi_0$ is any price that is high enough to guarantee a transaction (i.e. $\pi_0$ is effectively $\infty$).  We can then rewrite $p(x | s)$ without any restrictions on the range in the form
\begin{equation}
p(x | s) =  P_\theta (s) \delta(x - \pi_0) + H(x - s) p(x),
\label{decomp}
\end{equation} 
where $H(x)$ satisfies $H(x) = 1$ for $x < 0$ and $H(x) = 0$ for $x > 0$, and $\delta(x)$ is the Dirac delta function, which satisfies $\delta(x) = 0$ for $x \ne 0$ and $\int \delta(x) dx = 1$ for any range of integration that includes $0$.  

We can test our hypothesis by assuming it true and then testing whether this is consistent with the data. 
To do this we can get better statistical convergence by aggregating a semi-infinite range of values of the spread by studying $p(x | s > s_1)$, where $s_1$ is a threshold that can be varied to test for dependence on the spread.  This conditional density contains information about both the right and left tail of the distribution.  Since $s$ can be arbitrarily large, $p(x | s > s_1)$ contains limit orders with arbitrarily large values of $x$ .  However, the sample distribution will contain a different number of points on the left and right sides due to the fact that for $x > s_1$ the probability of $x$ depends on the distribution of the spread.  $p(x | s > s_1)$ can be written
\[
p(x | s > s_1) = \frac{\int_{s_1}^\infty p(x | s) p(s) ds}{\int_{s_1}^\infty p(s) ds}.
\]
From equation~\ref{decomp} this becomes
\begin{equation}
p(x | s > s_1) = \frac{\delta(x - \pi_0) \int_{s_1}^\infty P_\theta (s) p(s) ds \\
 + p(x) \int_{s_1}^\infty H(x - s) p(s) ds}{\int_{s_1}^\infty p(s) ds}.
 \label{integrals}
\end{equation}

There are three ranges of $x$ that must be treated separately.
\begin{itemize}
\item
$x < 0$ corresponds to limit orders placed inside the book.  In this range $p(x | s > s_1) = p(x)$.  This can be explicitly verified by doing the integral over $H(x - s)$ in equation~\ref{decomp} over $s$.
\item
$0 < x < s_1$ corresponds to limit orders placed inside the spread, in which case $p(x | s > s_1)$ is given by equation~\ref{aboveS1} below.
\item
$s_1 \le x$ corresponds to trading orders that generate immediate transactions, with a probability given by equation~\ref{hyp2}.
\end{itemize}

We now compute a simple expression for $p(x | s > s_1)$ for the second bullet above, i.e. the range where $s_1 < x < \pi_0$.  By making a change of variables the Heaviside function can be removed and the domain of integration adjusted to include only the non-zero range of the integrand.  This gives
\begin{equation}
p(x | s > s_1) = p(x) \frac{\int_{x}^\infty p(s) ds}{\int_{s_1}^\infty p(s) ds}.
\label{fullTest}
\end{equation}

This gives an easy way to test our hypothesis.   Letting $N(x > z)$ be the number of limit orders that satisfy the condition $x > z$, equation~\ref{fullTest} gives an expression for estimating $p(x)$ for $x > s_1$ as 
\begin{equation}
p(x) = p(x | s > s_1)\frac{\int_{s_1}^\infty p(s) ds}{\int_{x}^\infty p(s) ds} \approx \frac{N(s > s_1)}{N(s > x)}.
\label{aboveS1}
\end{equation}

\begin{figure}[ptb]
\begin{center}
\includegraphics[scale=0.5]{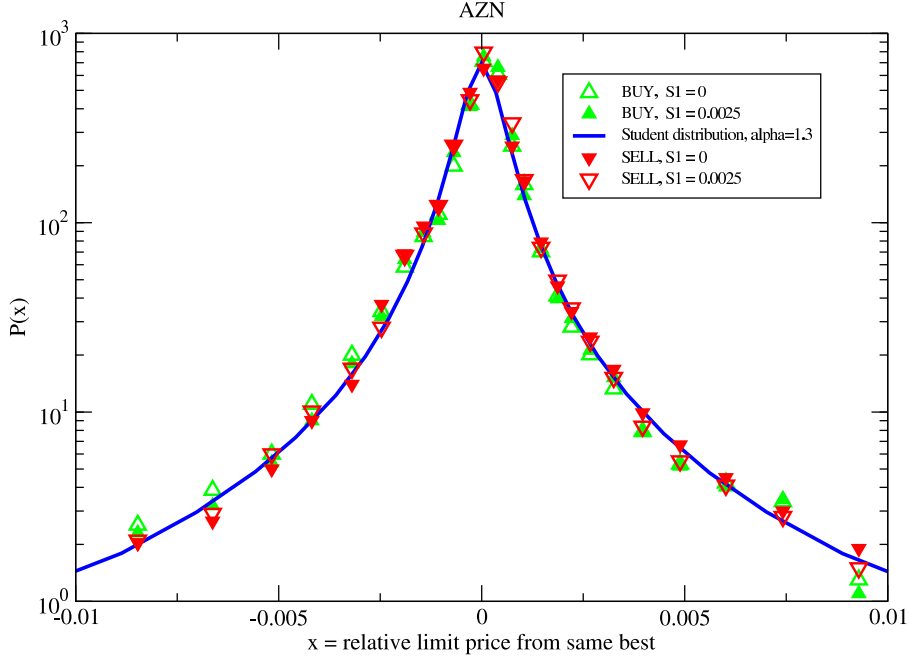}
\end{center}
\caption{Reconstruction of the probability density function $p(x)$ describing limit order prices as a function of $x$, the limit price relative to the same best price.  The reconstruction is done both for buy orders (green upward pointing triangles) and sell orders (red downward pointing triangles), and for two different values of the spread, $s_1 = 0$, which uses all the data, and $s_1 = 0.0025$, which uses roughly the $10\%$ of the data with the largest spread.  The fitted blue curve is a Student distribution with 1.3 degrees of freedom.  This shows the order placement is roughly symmetric, independent of the spread, and the same for buy and sell orders.}
\label{collapsePricePlot}
\end{figure}

Using these results we can reconstruct $p(x)$ for $-\infty < x < \infty$ with any spread condition $s_1$,  by using equation~\ref{aboveS1} for $x > s_1$ and $p(x | s > s_1)$ otherwise.   This is done for two different values of $s_1$ in Figure~\ref{collapsePricePlot}.  
The fact that the data collapse onto the same curve even though we use two very different spread conditions confirms that $p(x | s) \approx p(x)$.  This result is surprising for three reasons:
\begin{itemize}
\item
$p(x | s)$ is independent of the spread.  This is surprising because one would naively expect the strategic considerations of order placement to depend on the spread.
\item
The distributions for buying and selling are the same.
\item
The distribution is symmetric.  This is most surprising of all, since for reasons that we outlined in the introduction to this section, the strategic considerations for placing orders inside the spread seem quite different than those for placing an order inside the book.
\end{itemize}

\begin{figure}[ptb]
\begin{center}
\includegraphics[scale=0.5]{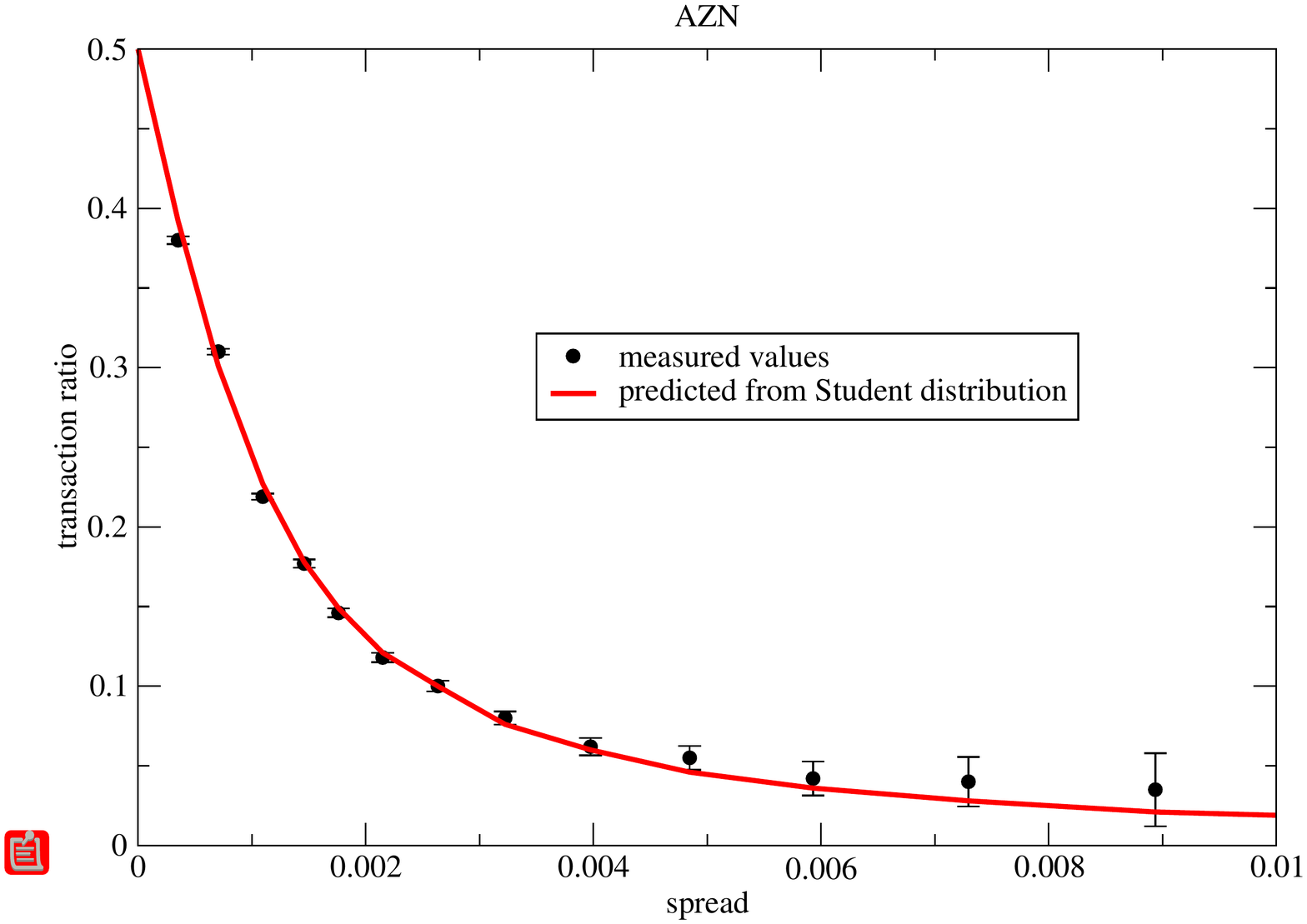}
\end{center}
\caption{The transaction probability $P_\theta$ as a function of the spread.  The curve is based on the fit to a student distribution for $p(x)$ in Figure~\ref{collapsePricePlot} and equation~\ref{hyp2}.  This demonstrates that the fraction of orders that result in transactions approaches $1/2$ in the limit as the spread goes to zero and approaches zero in the limit as the spread becomes large.}
\label{transactionRatio}
\end{figure}

We can test equation~\ref{hyp2} using the fit to the Student distribution from Figure~\ref{collapsePricePlot}.  In Figure~\ref{transactionRatio} we plot the fraction of orders that result in transactions as a function of the spread.  This gives an excellent fit to the data.  The probability that an order generates a transaction approaches $1/2$ in the limit as the spread goes to zero, and approaches zero in the limit as the spread becomes large.

To test whether this form is exact or is only an approximation, we used the Kolmogorov-Smirnov (KS) test.  In order to test the conjecture for different values of the spread, we divided the data set into bins based on the value of the spread, in the ranges shown in Table~\ref{ks}.  We then drew values from the Student distribution to match the number of points in each bin, and applied a two-sided KS test.  The statistics of the test for each bin and the corresponding $p$ values are given in Table~\ref{ks}. 
\begin{table}[htdp]
\begin{center}
\begin{tabular}{c|c|c|c|c|c|c|c|c|c}
s in units of $E[s]$ & 0 - 0.5 &  0.5 - 1 & 1 - 1.5 & 1.5 - 2 & 2 -3 & 3 - 5 & 5 - 10 & 10 - 20 \\
\hline
KS value & 0.39 & 0.89 & 1.12 & 0.97 & 1.2 & 1.07 & 1.32 & 1.15 \\
\hline
$p$ value & 0.72 & 0.2 & 0.08 & 0.15 & 0.05 & 0.1 & 0.03 & 0.07 \\
\end{tabular}
\caption{Results of applying the Kolmogorov-Smirnov test to the data shown in Figure~\ref{collapsePricePlot}.  The first row contains the range of spreads for each bin in units of the sample mean $E[s]$, the second row contains the values for the test, and the third row contains the corresponding $p$ values.}
\label{ks}
\end{center}
\end{table}
In one case out of eight we get a rejection at the $95\%$ confidence level, and in four cases out of eight we get rejections at the $90\%$ confidence level, suggesting that there are small deviations from the Student distribution hypothesis.  However, these rejections are all weak, and it is important to keep in mind that the KS test requires the data to be IID.  A close examination of the data makes it clear that this is not true.  Zovko and Farmer (\citeyear{Zovko02}) studied negative values of $x$, and found that they have a strong positive autocorrelation in the tail; we also observe that $|x|$ has significant autocorrelations.   Considering this the results of the KS test are surprisingly good.  In any case it is clear that the Student distribution is a very good approximation of the true distribution.

This raises the question of whether $p(x)$ can correspond to an economic equilibrium.   As we have already suggested, the obvious strategy for placing an order inside the spread is to choose a price one tick better than the current same best price.  But perhaps this is naive once one considers a finite tick size and the price priority that comes from placing an order first.  When the spread is wide it is likely that other limit orders will arrive soon and cause the impatient trader to move her order again.  When she places a new order she may forced to a price that is already occupied and thus get lower priority.  In this case she would have been better off to quote a more aggressive price to begin with. Under appropriate assumptions such a line of reasoning might result in an economic equilibrium.   Because the strategic considerations inside and outside the book are so different, however, it is hard to believe that such an equilibrium distribution would be symmetric around the best price.  This suggests that this result may only be explainable on behavioral grounds.  At this stage it is unclear whether we should consult an equilibrium theorist or a psychologist\footnote{Another alternative is that we should consult an expert in non-equilibrium statistical mechanics.  The Student distribution occurs as a standard solution in nonequilibrium statistical mechanics.   An alternate explanation might be an equilibrium that is constrained by inattention, e.g. because of time lags between price perception and order placement.  Note, though, that in the LSE order placement is essentially instantaneous and time stamps are accurate to within the second.}

\section{Long-memory and order signs \label{orderSigns}}

To model order placement it is necessary to decide whether each new order is to buy or to sell.  We arbitrarily designate $+1$ for buy and $-1$ for sell.  Given that returns are essentially uncorrelated in time, it might seem natural to simply assume that order signs are IID.  However, this is not a good approximation for the two markets where this has been studied\footnote{These studies were for the Paris and London stock markets.}.  Instead, the signs of orders follow a long-memory process (Bouchaud et al. \citeyear{Bouchaud04}, Lillo and Farmer \citeyear{Lillo03c}).  Roughly speaking, this means that the autocorrelation of order signs $C(\tau)$ is not integrable, and decays as $\tau^{-\gamma}$ for large $\tau$, $0 < \gamma < 1$.  Here $\tau$ is the time lag between the placement of two orders measured either as the intervening number of transactions or in terms of clock time while the market is open.  The coefficients of the sample autocorrelation remain positive at statistically significant levels for lags of $10,000$ transactions, corresponding to time intervals of several weeks.  This is surprising because it implies a high degree of predictability in order signs -- by observing the sign of an order that has just been placed, it is possible to make a statistically significant prediction about the sign of an order that will be placed two weeks later.   In order to compensate for this and keep price changes uncorrelated, the market must respond by adjusting other properties, such as liquidity, in order to compensate (Lillo and Farmer \citeyear{Lillo03c}).   As Bouchaud et al. (\citeyear{Bouchaud04}, \citeyear{Bouchaud04b}) have pointed out this also implies adjustments in subsequent order placement.  We find that the long-memory properties of order signs is very important for price formation.

We have offered a model to explain the long-memory of order flow based on strategic order splitting (Lillo, Mike, and Farmer \citeyear{Lillo05}). When an agent wishes to trade a large amount, she does not do so by placing a large trading order, but rather by splitting it into smaller pieces and executing each piece incrementally according to the available liquidity in the market.  We make a model in which we assume hidden orders have an asymptotic power law distribution in their size $V$ of the form $P(V > v) \sim v^{-\beta}$, with $\beta > 0$, as observed by Gopikrishnan et al. \citeyear{Gopikrishnan00}).  Our model assumes that hidden orders enter according to an IID process, and that they are executed in constant increments at a fixed rate.  We show that the signs of the executed orders are a long-memory process whose autocorrelation function has exponent $\gamma = \beta - 1$.  This prediction is borne out empirically by comparisons of off-book and on-book  data (Lillo, Mike and Farmer, \citeyear{Lillo05}).

The customary way to discuss long-memory is in terms of the Hurst exponent, which is related to the exponent of the autocorrelation function as $H = 1 - \gamma/2$.  For a long-memory process the Hurst exponent is in the range $1/2 < H < 1$.   For a diffusion process with long-memory increments the variance over a period $t$ scales as $t^{2H}$, and statistical averages converge as $t^{(H - 1)}$.  This creates problems for statistical testing, as discussed in Section~\ref{priceFormation}.

For simulating price formation as we will do in Section~\ref{priceFormation} we have used our model described above, and we have also used a fractional gaussian random process (Beran, \citeyear{Beran94}) (in the latter case we take the signs of the resulting random numbers).   Because the algorithm for the fractional gaussian algorithm is standard and easy to implement, for purposes of reproducibility we use it for the results presented here.  The studies cited above examined the signs of market orders only, but in our model we use a long-memory process to model the signs of market orders and limit orders.  This is justified by studies that we have done of the signs of limit orders, which we find exhibit long-memory that is essentially equivalent to that of market orders.

\section{Order cancellation \label{orderCancellation}}

Cancellation of trading orders plays an important role in price formation.   It causes changes in the midprice when the last order at the best price is removed, and can also have important indirect effects even when it occurs inside the limit order book.  It affects the distribution of orders in the limit order book, which can later affect price responses to new market orders.  Thus it plays an important role in determining liquidity.

The model that we develop for cancellation in this section is crude and should be regarded as preliminary.  Our reason for presenting our results at this time is that we want to illustrate that the order placement from the previous section plays an important role in price formation.  Although the analysis leading up to the cancellation model developed here is quick and dirty, it is justified in the end by the fact that it produces fairly good results.  The resulting model is simple and provides a basis for improvement.

\begin{figure}[ptb]
\begin{center}
\includegraphics[scale=0.5]{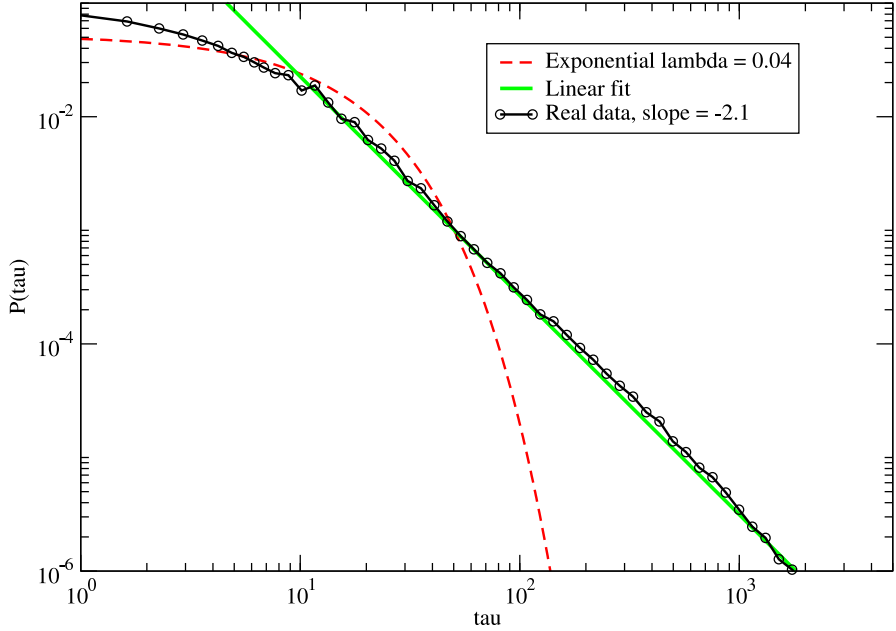}
\caption{The empirical probability density of the lifetime $\tau$ of cancelled orders for the stock Astrazeneca (black).  $\tau$ is the number of trading orders placed between the time given order is placed and the time it is cancelled. This is compared to an exponential distribution with $\lambda = 0.03$ (red).  A power law $\tau^{-(1 + \gamma)}$ with $\gamma = 1.1$ is shown for comparison.  Note that to avoid end of day effects we exclude orders that are are not cancelled between 9:00 am and 4:00 pm on trading days (but we do include orders that are placed on one day and cancelled on another day).}
\label{lifetimes}
\end{center}
\end{figure}

The zero intelligence model used the crude assumption that cancellation is a Poisson process.   Let $\tau$ be the lifetime of an order measured from when it is placed to when it is cancelled, where time is measured in terms of number of trading orders that are placed (either market orders or limit orders).
Under the Poisson assumption the distribution of lifetimes is an exponential distribution of the form $P(\tau) = \lambda (1 - \lambda)^{\tau - 1}$, where $\lambda$ is the rate of the Poisson process.  This can also be written $\lambda = 1/E[\tau]$, where $E[\tau]$ is the expected lifetime.  For AZN, for example, $\lambda \approx 0.040$.  A comparison of the exponential to the true distribution, as done in  Figure~\ref{lifetimes}, makes it clear that the
Poisson process is a poor assumption.  The tail of the empirical density function behaves like a power law\footnote{We find that this exponent is not very stable -- when we subsample the data its value varies somewhat in different periods.} of the form $\tau^{-(\gamma + 1)}$,  with $\gamma = 1.1$. This is a good approximation over roughly two orders of magnitude. The fact that the tail is a power law implies that the most long-lived orders last an order of magnitude longer than they would under the Poisson hypothesis.   Similar results are observed for BLT with $\gamma \approx 1.55$ and LLOY with $\gamma \approx 1.19$.  

It is clear that the Poisson process needs to be modified and by a process whose lifetime distribution has a heavy tail.  There are several effects that can cause this.  For example, if $\lambda(\tau)$ is a decreasing function, the longer an order exists without being cancelled, the less likely it is to be cancelled.  Alternatively, even if $\lambda(\tau)$ is constant in time, if it depends on the order $i$ this can generate heavy tails in the lifetime distribution of the whole population.  We find three different effects that influence the lifetimes of orders, and also influence other properties of the order book that are important for price formation.  These three factors are position in the order book relative to the best price, imbalance of buy and sell orders in the book, and the total number of orders.  We now explore each of these effects in turn.

\subsection{Position in the order book}

Strategic considerations dictate that position in the order book should be important in determining the cancellation rate.   Someone who places an order with a positive value of $x$ is likely to have a very different expected execution time than someone who places an order with a negative value of $x$.  If an order is placed at the best price then this implies that the trader is impatient and more likely to cancel the order quickly if it is not executed.   In contrast, no one would place an order deep inside the book unless they are prepared to wait a long time for execution.  Dependence on cancellation times with these basic characteristics was observed in the Paris Stock Market by Potters and Bouchaud (\citeyear{Potters03}).

To study this effect we will measure the cancellation rate as a function of the distance to the opposite best price.  Letting $t_i$ be the time when the order was placed,  the distance from the opposite best at time $t - t_i > 0$ is $\Delta_i(t - t_i) = \pi - \pi_b(t)$ for sell orders and $\Delta_i(t) = \pi_a(t) - \pi$ for buy orders.  $\Delta(0)$ is thus the distance to the opposite best when the order is placed, and $\Delta(t) = 0$ if and when the order is executed.  We compute the sample correlation $\rho(\Delta(0),\tau)$, and find that $\rho \approx 0.23$ for AZN, $\rho \approx 0.14$ for BLT, and $\rho \approx 0.18$ for LLOY, confirming the positive association between distance to the opposite best and cancellation time.

Strategic considerations also suggest that cancellation should depend on $\Delta (t)$ as well as $\Delta(0)$.   If $\Delta(t) \gg \Delta(0)$ then this means that the opposite best price is now much further away than when the order was originally placed, making execution unlikely and making it more likely that the order will be cancelled.  Similarly, if $\Delta(t) \ll \Delta(0)$ the opposite best price is quite close, execution is very likely and hence cancellation should be less likely.  This is confirmed by fact that for buy cancellations we observe positive correlations with the opposite best price movements in the range of $20 - 25\%$, and for sell orders we observe negative correlations of the same size.  In the interest of keeping the model as simple as possible we define a variable that encompasses both the dependence on $\Delta(0)$ and the dependence on $\Delta(t)$, defined as their ratio
\[
y_i(t) = \frac{\Delta_i(t)}{\Delta_i(0)}.
\]
By definition when order $i$ is placed $y_i = 1$ and if and when it is executed, $y_i = 0$.  A change in $y_i(t)$ indicates a movement in the opposite price in units whose scale is determined by where order $i$ was originally placed.  

\begin{figure}[ptb]
\begin{center}
\includegraphics[scale=0.4]{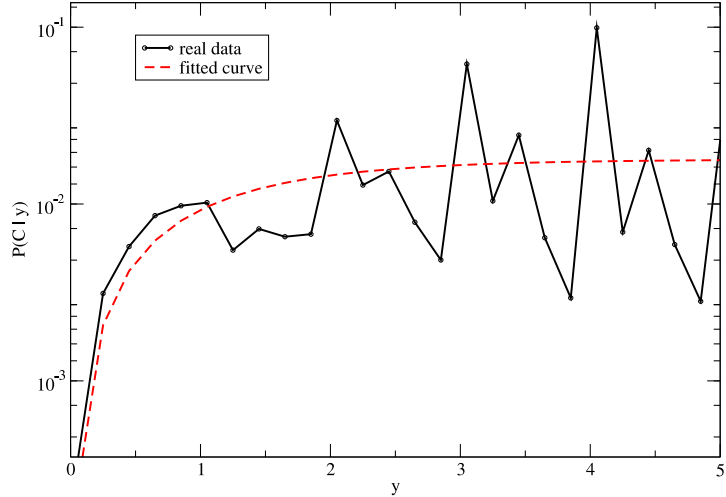}
\caption{The probability of cancellation $P(C | y_i)$ for AZN conditioned on $y_i(t) = \Delta_i(t)/\Delta_i(0)$.  The variable $y_i$ measures the distance from order $i$ to the opposite best price relative to its value when the order was originally placed.  The solid curve is the empirical fit $K_1(1-e^{-x})$, with $K_1 \approx 0.012$.}
\label{yProb}
\end{center}
\end{figure}

To measure the conditional probability of cancellation we use Bayes' rule.  The probability of cancellation conditioned on $y_i$ can be written 
\begin{equation}
P(C | y_i) = \frac{P(y_i | C)}{P(y_i)} P(C),
\label{bayes}
\end{equation}
where $C$ is a variable that is true when a cancellation occurs and false otherwise.  The conditional probability $P(y_i | C)$ can be computed by simply making a histogram of the values of $y_i$ when cancellations occur.  Figure~\ref{yProb} shows an empirical estimate of the conditional probability of cancellation for AZN computed in this way.  
Although there are substantial oscillations\footnote{We believe these oscillations are caused by round number effects in order placement and cancellation.}, as predicted by strategic considerations the cancellation probability tends to increase with $y_i$.  As $y_i$ goes to zero the cancellation probability also goes to zero, and it increases to a constant value of roughly $3\%$ per unit time as $y_i$ gets large (we are measuring time in units of the number of trading orders that are placed).  To approximate this behavior for modeling purposes we empirically fit a function of the form $K_1(1 - \exp(-x))$.  For AZN minimizing least squares gives $K_1 \approx 0.012$. 

The question remains whether the ratio $\Delta_i (t)/\Delta _i (0)$ fully captures the cancellation rate, or whether the numerator and denominator have separate effects that are not well modeled by the ratio.  To test this we divided the data into four different bins according to $\Delta_i (0)$ and repeated the measurement of Figure~\ref{yProb} for each of them separately.  We do not get a perfect collapse of the data onto a single curve.  Nonetheless, each of the four curves has a similar shape, and they are close enough that in the interest of keeping the model simple we have decided not to model these effects separately. 

\subsection{Order book imbalance} 

\begin{figure}[ptb]
\begin{center}
\includegraphics[scale=0.5]{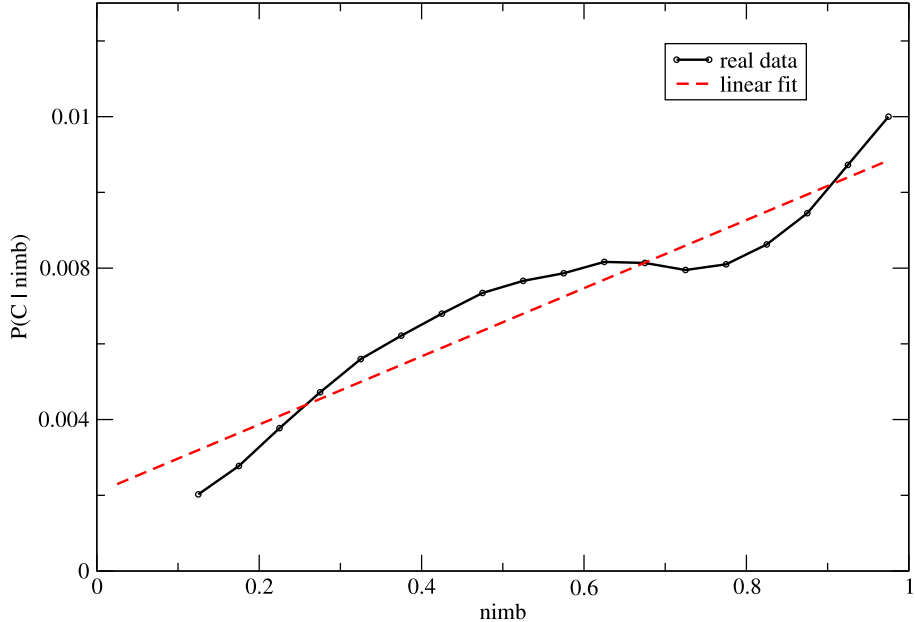}
\caption{The probability of cancellation per order, $P(C | n_{imb})$ for the stock AZN.  This is conditioned on the order imbalance $n_{imb}$.  The dashed curve is a least squares fit to a linear function, $K_2(n_{imb} + B)$, with $K_2 \approx 0.098$ and $B \approx 0.20$.}
\label{nimb}
\end{center}
\end{figure}

The imbalance in the order book is another factor that has a significant effect on order cancellation.   We define an indicator of order imbalance for buy orders as $n_{imb} = n_{buy}/(n_{buy} + n_{sell}) = $ and for sell orders as $n_{imb} = n_{sell}/(n_{buy} + n_{sell})$, where $n_{buy}$ is the number of buy orders in the limit order book and $n_{sell}$ is the number of sell orders.  In Figure~\ref{nimb} we show an empirical estimate of the conditional distribution $P(C | n_{imb})$, defined as the probability of cancellation per order.
$P(C | n_{imb})$ is less than $1\%$ when $n_{imb} = 0.1$ and about $4\%$ when $n_{imb} = 0.95$, increasing by more than a factor of four.  This says that it is more likely for an order to be cancelled when it is the dominant order type on the book, e.g. if the book has many more buy orders than sell orders, the probability that a given buy order will be cancelled increases (and the probability for a given sell order to be cancelled decreases).  Since the functional form appears to be a bit complicated, as a crude approximation we fit a linear function of the form $P(C | n_{imb}) = K_2( n_{imb} + B)$.  Minimizing least squares gives $K_2 \approx 0.098$ and $b \approx 0.20$ for AZN.

\subsection{Number of orders in the order book}

\begin{figure}[ptb]
\begin{center}
\includegraphics[scale=0.5]{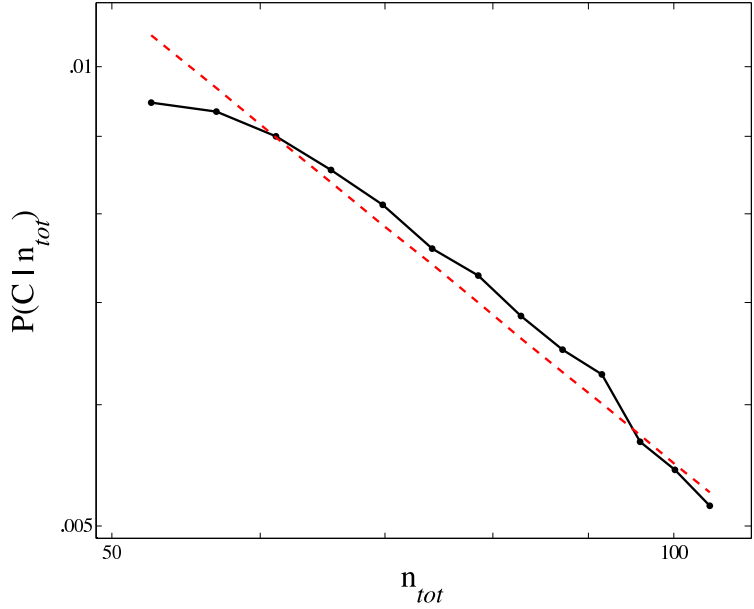}
\end{center}
\caption{The probability of cancellation per order, $P(C | n_{tot})$, for the stock AZN, conditioned on the total number of orders in the order book, $n_{tot}$ on a log-log plot.  The dashed line is the function $K_3/n_{tot}$, shown for reference, where $K_3=72.2$.\label{ntot}}
\end{figure}

Another variable that we find has an important effect on cancellation is $n_{tot}$, the total number of orders in the order book.  Using a procedure similar to those for the other two variables, in Figure~\ref{ntot} we plot the cancellation probability as a function of $n_{tot}$.
Surprisingly, we see that the probability of cancellation decreases as $n_{tot}$ increases, approximately proportional to $1/n_{tot}$.  A least squares fit of $\log P(C | n_{tot})$ vs. $b - a \log n_{tot}$ gives a slope $a = 0.92 \pm 0.06$ (using one standard deviation error bars).  The coefficient $a$ is sufficiently close to one that we simply make the approximation in our model that $P(C | n_{tot}) \sim 1/n_{tot}$.  We plot a line of slope $-1$ in the figure to make the validity of this approximation clear.

This is very surprising, as it indicates that the total cancellation rate is essentially independent of the number of orders in the order book.  It raises the question of how the order book can be stable:  If the total cancellation rate does not increase as the number of orders in the book increases, what prevents the number of orders in the book from blowing up?  As we discussion in Section~\ref{priceFormation}, we find that the order book dynamics are stable, even with this term.  We believe that this is achieved by the order placement dynamics illustrated in Figure~\ref{transactionRatio}.  As $n_{tot}$ increases, the spread tends to decrease, which stimulates the submission of more market orders, which decreases $n_{tot}$.

\subsection{Combined cancellation model} 

We assume that the effects of $n_{imb}$, $y_i$, and $n_{tot}$ are independent, i.e. the conditional probability of cancellation per order is of the form
\begin{equation}
P(C | y_i, n_{imb}, n_{tot}) = \frac{P(y_i | C) P(n_{imb} | C) P(n_{tot} | C)}{P(y_i)P(n_{imb})P(n_{tot})} P(C) = A(1-\exp^{-y_i})(n_{imb}+ B)/n_{tot},
\label{cancModel}
\end{equation}
where for AZN $A \approx 0.45$ and $B \approx 0.20$.

\begin{figure}[ptb]
\begin{center}
\includegraphics[scale=0.5]{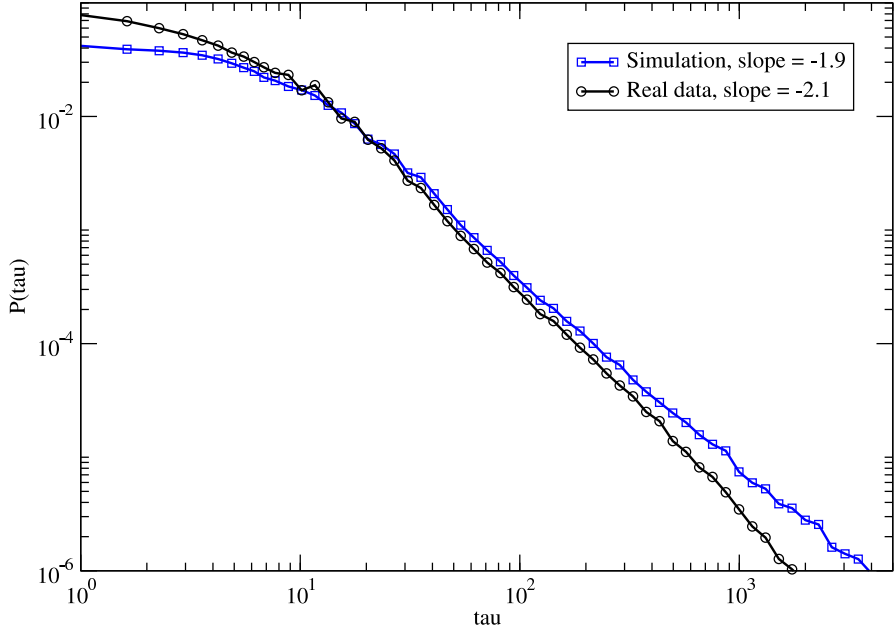}
\caption{A comparison of the distribution of lifetimes of simulated cancellations (blue squares) to those of true cancellations (black circles).}
\label{lifetimeComp}
\end{center}
\end{figure}

To test this model we simulate cancellations and compare to the real data.  Using the real data, after the placement of each new order we measure $y_i$, $n_{imb}$, and $n_{tot}$ and simulate cancellation according to the probability given by equation~\ref{cancModel}.   We compare the distribution of lifetimes from the simulation to those of the true distribution in Figure~\ref{lifetimeComp}. 
The simulated lifetime distribution is not perfect, but it is much closer to the true distribution than the Poisson model (compare to Figure~\ref{lifetimes}).  It reproduces the power law tail, though with $\gamma \approx 0.9$, in comparison to the true distribution, which has $\gamma \approx 1.1$.  For small values of $\tau$ the model underestimates the lifetime probability and for large values of $\tau$ it overestimates the probability.  As an additional test of the model we plotted the average number of simulated cancellations against the actual number of cancellation for blocks of 50 events, where an event is a limit order, market order, or cancellation.   As we would hope the result is close to the identity.  Since the resulting plot is uninteresting we do not show it here.

\section{Price formation\label{priceFormation}}

The models of order placement and order cancellation that we have developed here can be used to simulate price formation.   We make three additional simplifying assumptions.
\begin{itemize}
\item
{\it All limit orders have constant size}.  This is justified by our earlier study of the on-book market of the London Stock Exchange in (Farmer \citeyear{Farmer04} et al.).  There we showed that orders that remove more than the depth at the opposite best quote are rare.  Thus from the point of view of price formation we can neglect large orders that penetrate more than one price level in the limit order book, and simply assume that a transaction either removes all shares at the best quote or does nothing.  We define $\Pi$ as the fraction of time a transaction removes the best quote.  
\item
{\it Tick size}.  Tick size can have an important effect on prices.  For the purposes of this study we assume a constant logarithmic tick size of $3 \times 10^{-4}$.
\item
{\it Order placement time}.  The basic unit of time for the simulation and for the corresponding statistical analysis of the real data is an order placement (whether a market order or a limit order).   This assumption is justified by a recent study showing that fluctuations in transaction frequency play only a minor role in determining clustered volatility and the distribution of returns (Gillemot, Farmer, and Lillo, \citeyear{Gillemot05}).
\end{itemize}

The simulation of the model proceeds as follows.  Each time step corresponds to the generation of a trading order.   The order sign\footnote{Note that we are generating order signs exogenously.  This is justified, for example, under the assumption made in the model of Lillo, Mike, and Farmer (2005) that hidden order arrival is exogenous to price formation.}  is generated using a fractional gaussian process\footnote{In contrast to the more realistic model of order flow described in Section~\ref{orderSigns}, the fractional gaussian processes does not allow us to control the prefactor of the correlation function, but rather generates a constant prefactor $C \approx 0.15$.  We find that this is close enough to the true values in Table~\ref{paramSummary} that this does not make a difference.} as described in Section~\ref{orderSigns} with Hurst exponent $H_s$.  We then generate a trading order with that sign by drawing a relative limit price $x$ from a Student distribution with scale $\sigma$ and tail exponent $\alpha(x)$, based on the model developed in Section~\ref{orderPlacement}.  If $x < s$ we place a limit order at price $x$ relative to the opposite best, and otherwise we place a market order.   If the result is a market order then we remove all the orders at the opposite price with probability $\Pi$; otherwise the order has no effect.  We then determine which orders are cancelled by examining each order in the limit order book and generating random numbers with probability given by equation~\ref{cancModel} (which has parameters $A$ and $B$.  Note that more than one order can be cancelled in a given timestep.   The only exception is if this would result in either side of the limit order book becoming too empty.  To prevent this from happening we impose an additional condition on order cancellation, and require that after cancellation there be at least two orders remaining on the corresponding side of the book.  As a summary, in Table~\ref{paramSummary} we give the measured values of the seven parameters of the model for the three stocks in our data set.
\begin{table}[htdp]
\begin{center}
\begin{tabular}{c|c|c|c|c|c|c|c}
stock ticker & H & C & $\alpha(x)$ & $\sigma \times10^{-3}$ & $\Pi$ & A & B \\
\hline
AZN & 0.77 & 0.15 & 1.31 & $2.4$ & 0.45 & 0.45 & 0.20 \\
\hline
BLT & 0.80 & 0.18 & 1.55 & $2.0$ & 0.40 & 0.51 & 0.21 \\
\hline
LLOY & 0.81 & 0.13 & 1.25 & $2.6$ & 0.43 & 0.47 & 0.22
\end{tabular}
\caption{The measured parameters of our model for the three stocks in our data set.  H is the Hurst exponent of the order sign series, C is the prefactor of the sign autocorrelation function, $\alpha(x)$ and $\sigma$ are the tail exponent and the scale parameter of the order placement distribution, and $\Pi$ is the market order penetration rate.  The probability of cancellation for a given order is $P(C | y_i, n_{imb}, n_{tot}) = A(1-e^{-y_i})(n_{imb}+B)/n_{tot}$.\label{paramSummary}}
\end{center}
\end{table}

The parameters of the simulation for a given stock are based on the parameters in Table~\ref{paramSummary}.   The simulation generates a series of events that we compare to the real data in the same period where the parameters are measured. The particular sequence of events generated in this manner depends on the random number seed used in the simulation, and will obviously not match the actual data in detail.  We instead test the model by comparing its statistical properties to those of the real data.

\begin{figure}[ptb]
\begin{center}
\includegraphics[scale=0.4]{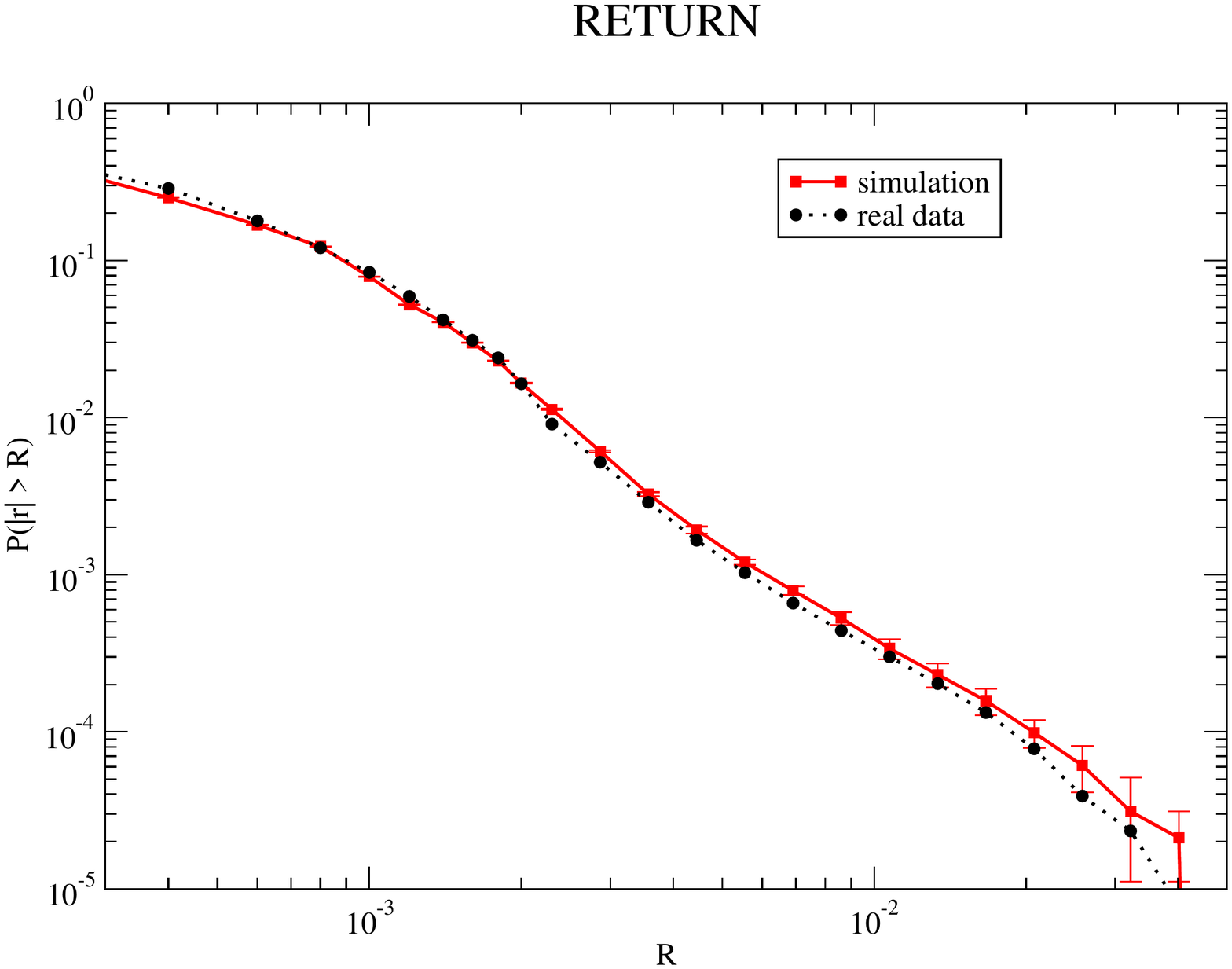}
\includegraphics[scale=0.4]{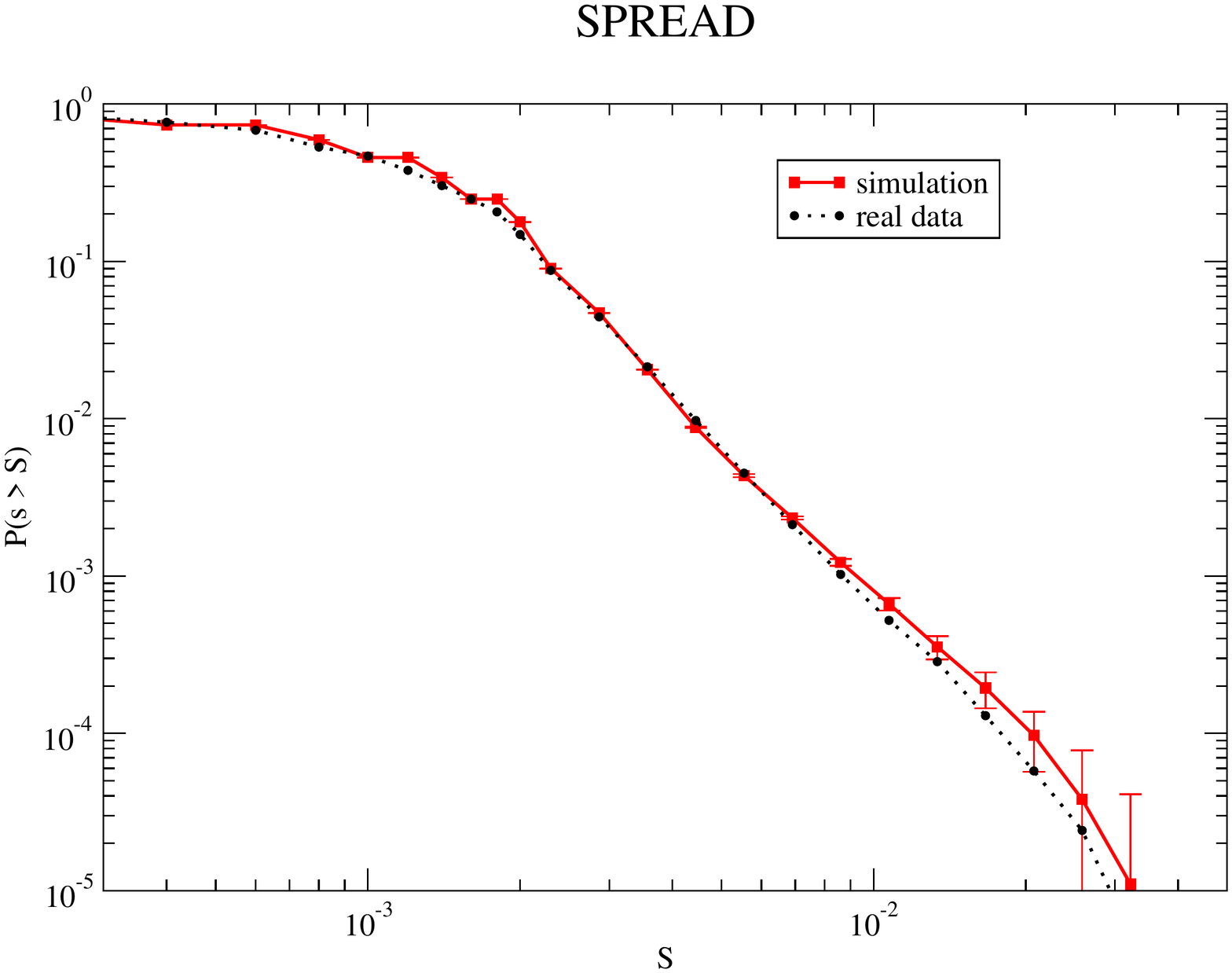}
\end{center}
\caption{A comparison of the distribution of absolute returns $|r|$ (upper) and spreads $s$ (lower) predicted by the model to those of the real data for the stock Astrazeneca.  The error bars for the model were computed by running the model for 20 times the number of events in the data set and taking the standard deviation over the 20 resulting distributions.  The solid curve is based on a single run of the model of length equal to the data set.} 
\label{returnDist}
\end{figure}

In Figure~\ref{returnDist} we compare the predictions of the model to the empirical distribution of returns and spreads for the stock AZN.
The returns are logarithmic midprice changes between transactions, i.e. $r(t_i) = \log p(t_i) - p(t_i - t_{i-1})$, where $t_i$ is the time when a transaction occurs and $p(t_i)$ is the midprice $p(t_i) = 1/2(\pi_a (t_i) + \pi_b(t_i))$ immediately after the transaction.  The distribution for the spread is built by recording the best bid and ask prices immediately before transactions\footnote{The time when the spread is recorded makes a significant difference in the distribution.  The spread tends to narrow after receipt of limit orders and tends to widen after market orders or cancellations.}.  The agreement is extremely good, both in terms of magnitude and functional form.

We want to stress that these predictions are made without any adjustment of parameters based on prices.  All the parameters of the model are based on the order flow process alone -- there are no adjustable parameters to match the scale of the returns or the spreads.  In Table~\ref{comparison} we compare summary statistics for the real data to those of the simulation for the three stocks in our sample.  We do this for the mean, the standard deviation, and the tail exponent of the spread and the transaction to transaction returns for each stock.  
\begin{table}[htdp]
\begin{center}
\begin{tabular}{|l|l|l|l|l|l|ll||}
\hline
stock ticker&  $E(|r|) \times 10^{-4}$  & $E(s) \times 10^{-4} $ &  $\sigma(|r|)\times 10^{-4}$ & $\sigma(s) \times 10^{-4}$ & $\alpha(|r|)$ & $\alpha(s)$ \\
\hline
AZN & 5.4 $\pm$ 1.2 & 13.9 $\pm$ 0.6 & 7.2 $\pm$ 2.1 & 12.1 $\pm$ 0.7 & 2.4 $\pm$ 0.2 & 3.3 $\pm$ 0.3\\
AZN predicted & 5.1 $\pm$ 0.1 & 14.3 $\pm$ 0.2 & 7.2 $\pm$ 0.3 & 13.2 $\pm$ 0.2 & 2.1 $\pm$ 0.4 & 3.0 $\pm$ 0.3\\
\hline
\hline 
BLT & 10 $\pm$ 0.9 & 24.1 $\pm$ 0.9 & 17.7 $\pm$ 1.6 & 27.1 $\pm$ 1.2 & 1.9 $\pm$ 0.2 & 2.8 $\pm$ 0.3\\
BLT predicted & 9.4 $\pm$ 0.1 & 22.7 $\pm$ 0.1 & 16.6 $\pm$ 0.4  & 26.4 $\pm$ 0.3 & 2.2 $\pm$ 0.4 & 2.7 $\pm$ 0.3\\
\hline
\hline
LLOY & 7.6 $\pm$ 1.3 & 17.1 $\pm$ 0.9 & 8.9 $\pm$ 0.6 & 13.8 $\pm$ 0.5 & 2.4 $\pm$ 0.4 & 3.5 $\pm$ 0.2\\
LLOY predicted & 7.4 $\pm$ 0.1& 16.8 $\pm$ 0.1 & 9.2 $\pm$ 0.1 & 12.9 $\pm$ 0.3 & 2.2 $\pm$ 0.3 & 3.2 $\pm$ 0.2\\
\hline
 \end{tabular}
\caption{Comparing the model with real data for three stocks in our sample.  $E(x) = $ sample mean of $x$, $\sigma(x) =$ standard deviation of $x$, and $\alpha(x) = $ tail exponent of $x$.  These statistics are based on the absolute logarithmic returns $|r|$ between transactions and the spread $s$ sampled as described in the text.  Error bars are one standard deviation, computed using the variance plot method.\label{comparison}}
\end{center}
\end{table}
The agreement is quite good across the board.  Out of $3 \times 6 = 18$ possible cases, in all but two cases the one standard deviation error of the prediction is within the one standard deviation error of the real value.  The error in the prediction is in every case less than $10\%$, and in most cases less than $5\%$.  

The reason the error bars in Table~\ref{comparison} seem large is because both the absolute returns and the spreads are long-memory processes.  Long-memory has been reported in volatility by many authors (Breidt, Crato and de Lima \citeyear{Breidt93}, Harvey \citeyear{Harvey93}, Ding, Granger and Engle \citeyear{Ding93}, Bollerslev and Mikkelsen, Baillie \citeyear{Baillie96}, \citeyear{Bollerslev96}, Willinger, Taqqu, and Teverovsky, \citeyear{Willinger99}, Gillemot, Farmer and Lillo \citeyear{Gillemot05}).  We confirm this here, getting $H = 0.78$ for AZN, $H = 0.74$ for BLT, and $H = 0.78$ for  LLOY.  As far as we know we are the first to report long-memory in spreads;  sampling before transactions, we find $H = 0.81$ for AZN, $H = 0.86$ for BLT, and $H = 0.79$ for LLOY\footnote{Note that both the tails and the autocorrelations of the spread depend strongly on sampling, due to the interaction between transactions and the spread.}.  While we have not performed significance tests (there is no good test that we know of), given the large size of our data set and the fact that these are so large in every case, we are confident that these numbers are all greater than $H = 0.5$ by a statistically significant amount, and these are long-memory processes.

For long-memory processes the statistical error scales as $N^{(H - 1)}$, where $N$ is the number of data points and $H$ is the Hurst exponent.   Standard methods for estimating errors fail when applied to long-memory processes, and the only applicable method that we are aware of is the variance plot method (Beran \citeyear{Beran94}).  The data set of length $N$ is divided into $m$ disjoint regions of length $L = \text{int} (N/m)$, where $\text{int}(x)$ is the integer part of $x$, and any remaining data are discarded.  The statistic of interest (the mean $E$, the standard deviation $\sigma$ or the tail exponent $\alpha$) is computed for each region based on the $L$ data points in that region.  The resulting $m$ values are used to compute the standard deviation $\Sigma(m)$ of the statistic across the $m$ regions.  This is done for a variety of different values of $m$.  Then $\log \Sigma(m)$ is regressed against $\log m$ assuming the linear relation $\log \Sigma(m) = (1 - H) \log m + b$, where $b$ is a free parameter.  We use values of $H$ computed using the DFA method based on polynomials of degree one (Deng et al. \citeyear{Peng94}).  The result is extrapolated to $m = 1$ to compute the error corresponding to the $p$ value for one standard deviation.

We have not yet had the opportunity to fully explore the properties of this model.   A fuller investigation would test the model on data from many different stocks and on more properties of the price series.   The model has several defects that we know of.  In particular, the autocorrelation function of returns drops to zero slower than the real data (it takes on values of the order of $1\%$ for about 50 time steps).  Another difference concerns clustered volatility.   While the model displays some clustered volatility, it is weaker and less persistent than that of the real data.  For example, for AZN the Hurst exponent of volatility of the model is $H = 0.64$, in contrast to $H = 0.78$ for the real data.   Another feature of this model that we do not like is that the {\it ad hoc} requirement that we preserve at least two orders in each side of the limit order book makes a noticeable difference in the statistical properties of the simulation, indicating that our existing cancellation model has not fully captured the order book dynamics.

Despite these caveats, the model does an extremely good job of describing the distribution of both returns and spreads.   The prediction of the tail behavior for both returns and spread is accurate to within statistical error.  Perhaps even more important, unlike previous models, the prediction is for the entire distribution, and not just for the tail exponent.  Although we cannot make a firm statement based on only three cases, the predicted tail exponent tracks the variations in the real tail exponents of the three stocks we study here quite well.  Though not yet fully convincing, this suggests that the tail exponents are not universal, but rather vary from stock to stock\footnote{Note also that all three values of $\alpha(|r|)$ are more than two standard deviations below the value of $3$ that is predicted by Gabaix et al. (\citeyear{Gabaix03}).}.  Our model may thus be useful for understanding the factors that cause variations in the behavior of extreme risk between stocks.  

\begin{figure}[tb]
\begin{center}
\includegraphics[scale=0.4]{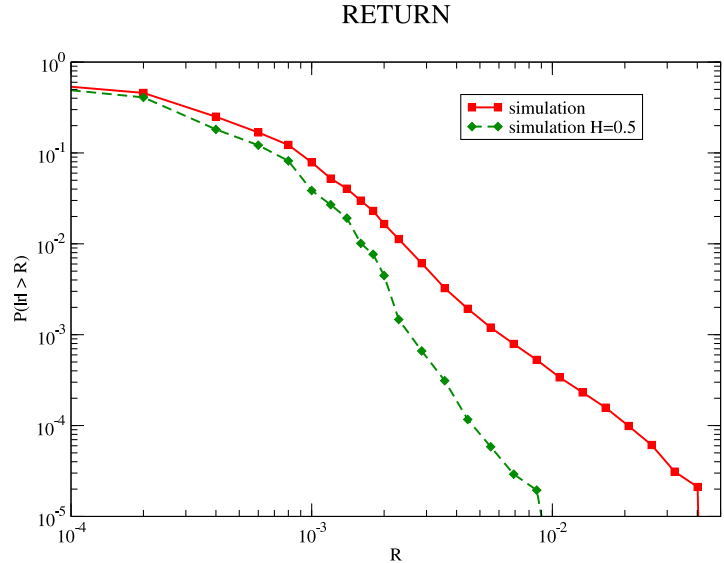}
\end{center}
\caption{A comparison of the distribution of absolute returns $|r|$ for the model using a long-memory sign generation process (red squares) and an IID sign generation process (green diamonds).  All other parameters are held fixed.} 
\label{returnComparison}
\end{figure}

Although we have not had a chance to make an exhaustive study of the behavior of the model under variation of its parameters, certain features are clear.  For example, the long-memory in the order signs is an important driver of the heavy tails of both the return and the spread distributions.  To demonstrate this, in Figure~\ref{returnComparison} we turn off the long-memory of order signs and replace it with IID process with a $50\%$ chance of generating a buy order and a $50\%$ chance of generating a sell order.    
Although still not normally distributed, the tails of the return distribution are clearly much thinner.   The tail exponents of returns $\alpha(|r|)$ and spreads $\alpha(s)$ also depend on other properties of the model, such as the tail exponent for limit order placement $\alpha(x)$ and parameters of the cancellation process, as well as the Hurst exponent $H$ of the order sign process.  The return distribution is not just a simple reflection of the order placement process, but involves an interaction between all the parts of the model that we do not yet fully understand.
       
\section{Conclusions}

The results that we have presented here are useful for several reasons.  On a practical level they give insight into the factors that generate financial risk.  They suggest that market microstructure is important.  The distribution of returns in the model is determined by the interaction of order placement and cancellation.  According to our model the heavy tails in price returns are driven by the long-memory of order signs, but they are also influenced by order placement and cancellation processes that also have heavy tails.  This prediction is quite different from previous models that say that they are due to large transactions (Gabaix et al, \citeyear{Gabaix03}), or that they are caused in a more generic way by nonlinear market dynamics, as discussed in the introduction.  Resolving this question is interesting for market design because it suggests that by altering properties of the order placement and cancellation process (e.g. by providing incentives to change behavior) it may be possible to alter the properties of financial risk.

These results are also interesting at the broader level of methodology.  It has become traditional in economics to require that all theories begin by modeling preferences.  A typical example is selfish utility maximization under rational choice.  At the other end are econometric models, which fit functional forms to data using purely {\it ad hoc} assumptions.  Our model is somewhere in between.  We make an econometric model for order placement and use this as the foundation to make predictions about price formation.  The model of order placement gives insight into decision making, but does not make any fundamental assumptions about the motivations that drive it.  This can be regarded as a divide and conquer strategy:  Rather than trying to start from first principles and work forward, we search for regularities in the middle.  From here one can either work forward to prices or work backward to try to discover what preferences and strategic considerations might generate these regularities.  We have chosen to take the much easier path of working forward because it leads to accurate quantitative predictions via a tractable research program.  

The path backward is also very interesting, but much less straightforward.  We cannot claim to understand the behavior we have documented here until we understand its motivation.   Nonetheless, by looking at order placement rather than directly at prices we are able to capture intermediate behavioral regularities that provide important clues about their origins, providing a useful benchmark for any attempt to make more fundamental models that address preferences and choice.  Is it possible to explain these regularities in terms of rational choice?  Or do they represent an example of irrational behavior, that can only be explained in terms of human psychology?

Finally our results are interesting because of the accuracy of their predictions. Based on properties of order flow, we are able to predict both the magnitude and the functional form of the return distribution to a high degree of accuracy.  The predictions of this model have the kind of accuracy that one expects in physics.  The research program is in the spirit of typical models in physics, which connect a set of empirical assumptions at one level to empirical phenomena at another level, but typically do not attempt to derive results from first principles.    

\acknowledgments We would like to thank the James S. McDonnell
Foundation for their Studying Complex Systems Research Award, Credit
Suisse First Boston, Bob Maxfield, and Bill
Miller for supporting this research. We would like to thank Fabrizio Lillo, Constantino Tsallis, Laszlo Gillemot and J-P. Bouchaud for useful discussions, and Marcus Daniels for technical support.  We would particularly like to thank Austin Gerig for reproducing many of these results and for providing Figure~\ref{ntot}.

\bibliographystyle{abbrvnat}
\bibliography{jdf} 

\end{document}